\documentclass[a4paper,12pt]{article}
\usepackage{graphics}
\usepackage{amsfonts}

\begin{document}
%
%\addtolength{\baselineskip}{.7mm}

\newcommand{\be}{\begin{equation}}
\newcommand{\ee}{\end{equation}}
\newcommand{\bea}{\begin{eqnarray}}
\newcommand{\eea}{\end{eqnarray}}
\newcommand{\bean}{\begin{eqnarray*}}
\newcommand{\eean}{\end{eqnarray*}}
\font\upright=cmu10 scaled\magstep1
\font\sans=cmss12
\newcommand{\ssf}{\sans}
\newcommand{\stroke}{\vrule height8pt width0.4pt depth-0.1pt}
\newcommand{\Z}{\hbox{\upright\rlap{\ssf Z}\kern 2.7pt {\ssf Z}}}
\newcommand{\ZZ}{\Z\hskip -10pt \Z_2}
\newcommand{\C}{{\rlap{\upright\rlap{C}\kern 3.8pt\stroke}\phantom{C}}}
\newcommand{\R}{\hbox{\upright\rlap{I}\kern 1.7pt R}}
\newcommand{\HH}{\hbox{\upright\rlap{I}\kern 1.7pt H}}
\newcommand{\CP}{\hbox{\C{\upright\rlap{I}\kern 1.5pt P}}}
\newcommand{\identity}{{\upright\rlap{1}\kern 2.0pt 1}}
\newcommand{\half}{\frac{1}{2}}
\newcommand{\quart}{\frac{1}{4}}
\newcommand{\pr}{\partial}
\newcommand{\bm}{\boldmath}
\newcommand{\I}{{\cal I}} 
\newcommand{\M}{{\cal M}}
\newcommand{\N}{{\cal N}}
\newcommand{\e}{\varepsilon}

\thispagestyle{empty}
\rightline{DAMTP-2018-34}
\vskip 3em
\begin{center}
{{\bf \large Force between Kinks with Long-range Tails
}} 
\\[15mm]

{\bf \large N.~S. Manton\footnote{email: N.S.Manton@damtp.cam.ac.uk}} \\[20pt]

\vskip 1em
{\it 
Department of Applied Mathematics and Theoretical Physics,\\
University of Cambridge, \\
Wilberforce Road, Cambridge CB3 0WA, U.K.}
\vspace{12mm}

%%%%%%%%%%%%%%%%%%%%%%%%%%%%%%%%%%%%%%%%%%%%%%%%%%
\abstract
{In a scalar field theory that has a symmetric octic potential with a
quartic minimum and two quadratic minima, kink and mirror kink 
solutions have long-range tails. We calculate the force between these 
kinks when their long-range tails overlap. This is
a nonlinear problem, solved using an adiabatic ansatz for the 
accelerating kinks that leads to a modified, first-order Bogomolny 
equation. We find the force is repulsive and decays with the fourth 
power of the kink separation.}
%%%%%%%%%%%%%%%%%%%%%%%%%%%%%%%%%%%%%%%%%%%%%%%%%%

\end{center}

\vskip 150pt
\leftline{Keywords: Kinks, Long-range tail, Scalar field, Force}
%\leftline{PACS: }
\vskip 1em

\vfill
\newpage
\setcounter{page}{1}
\renewcommand{\thefootnote}{\arabic{footnote}}

%%%%%%%%%%%%%%%%%%%%%%%%%%%%%%%%%%%%%%%%%%%%%%%%%%%%%%%%%%%%%%%%%%%%%%%%%%%%%
%%%%%%%%%%%%%%%%%%%%%%%%%%%%%%%%%%%%%%%%%%%%%%%%%%%%%%%%%%%%%%%%%%%%%%%%%%%%%

\section{Introduction} 
\vspace{4mm}

Consider a Lorentz-invariant scalar field theory in $1+1$ dimensions
with Lagrangian density
\be
{\cal L} = \half \left(\frac{\pr\phi}{\pr t}\right)^2 
- \half\left(\frac{\pr\phi}{\pr x}\right)^2 - V(\phi) \,.
\label{Lagran}
\ee
Its field equation is
\be
\frac{\pr^2\phi}{\pr t^2} - \frac{\pr^2\phi}{\pr x^2} +
\frac{dV}{d\phi} = 0 \,.
\label{fieldeq}
\ee
Assume that the potential $V(\phi)$ is smooth and 
non-negative, and attains its
minimum value $V=0$ for one or more values of $\phi$. These field
values are vacua of the theory, and if $V$ has more than one vacuum, 
then it has static kink solutions interpolating between them as $x$ 
increases from $-\infty$ to $\infty$. Much is known about kinks 
for generic $V$, and also in particular examples \cite{book,Shn}. 
The classic kink occurs for the quartic potential $V(\phi) = 
\quart(1 - \phi^2)^2$. The solution $\phi(x) = \tanh(x)$ 
connects the vacua $\phi = -1$ and $\phi = 1$, and there is an
antikink going the other way. If the theory is 
extended to higher spatial dimensions, then the kink becomes a domain 
wall, but we will not consider this generalisation further.

If $V$ has a finite or infinite sequence
of vacua $\phi_n$ (in increasing order) then there is a
kink connecting $\phi_n$ to $\phi_{n+1}$ and an antikink connecting 
$\phi_{n+1}$ to $\phi_n$ for each $n$, but there are no kinks
connecting vacua whose indices differ by more than one. Field 
configurations connecting non-adjacent vacua can be interpreted 
as nonlinear superpositions of more than one kink, but these
configurations are never static. Physically, the kinks repel
each other and separate.

The field equation for a static kink is a second-order ODE, but it can
generally be reduced to a first-order equation. This is a special case
of the Bogomolny trick \cite{Bo} (although known much earlier in this 
context). We shall refer to the first-order equation as the 
Bogomolny equation for the kink, and use it repeatedly, even 
in the context of kink dynamics.

The minima of a generic potential are all quadratic, 
with $V$ having positive second derivative. In this case
the tails of kinks are short-ranged, i.e. the field $\phi$ approaches
the vacuum values, between which it is interpolating, exponentially 
fast as $x \to \pm\infty$. However, Lohe \cite{Loh}, and more recently 
Khare et al. \cite{KCS} and Bazeia et al. \cite{BMM}, have
drawn attention to several examples where at least one minimum is not
quadratic. This is not generic, but easily occurs as parameters in $V$
are varied. If one of the minima of $V$ is quartic (the next simplest case)
then one tail of the kink that approaches it is long-ranged; $\phi$ 
approaches the quartic minimum with a $\frac{1}{x}$ behaviour.

When the tails are short-ranged, then it is possible to calculate the
force between two well-separated kinks \cite{book}. 
This is for the situation where the kink on the left (smaller $x$) 
interpolates between $\phi_{n-1}$ and $\phi_n$, and the kink on the 
right interpolates between $\phi_n$ and $\phi_{n+1}$. The calculation
relies on a linear superposition of the exponentially small tails in
the region between the kinks. The result is a force that
decays exponentially fast with the kink separation. An example is 
discussed in Appendix B.

There are a number of ways of approaching the force calculation. 
The force on one of the kinks can be found using a version of 
Noether's theorem to determine the rate of
change of its momentum \cite{Ma5}. This is equivalent to using the
energy-momentum tensor to find the stress exerted on the half-line
containing the kink. An alternative is to attempt to
approximately solve the full, time-dependent field equation. One can
make an ansatz describing an accelerating kink, and then match the
tail of this to the tail of the other
kink \cite{Man}. This determines the acceleration. A cruder
approach is simply to set up a static field configuration that
incorporates both kinks, satisfying the appropriate boundary conditions,
and estimate its energy as a function of the separation. The
derivative of this energy with respect to separation should give the
kink-kink force.

None of these methods is completely straightforward to implement, as
each depends on an ansatz for an interpolating field. The methods might not
agree. One also needs to know where the centres of the kinks are, to have a
precise notion of separation. The centres have to be carefully defined,
especially for kinks with an asymmetric profile, and we shall clarify
in Appendix A what a good definition is. Because of the extended 
character of kinks, any formula for the force usually makes sense 
only to leading order in the separation, even when the separation is
large, and subleading terms are meaningless.

The force between two kinks with long-range, $\frac{1}{x}$ tails has apparently 
not been accurately determined. Certainly, the calculation in \cite{book}
breaks down in this case. The force between a kink and antikink with
these tails was estimated to decay with the fourth power of the separation by
Gonz\'alez and Estrad-Sarlabous \cite{GE,MGGL}. The main purpose of
this paper is to establish a similar result for the kink-kink case,
and to find the numerical coefficient. We shall apply the different 
methods outlined above, and show to what extent they give consistent 
results. Our results show unambiguously that the force decays with the 
fourth power of the kink separation, but the coefficient is still not 
quite certain.  

Using the calculated force, we can find an effective equation of
motion for the positions of the kinks. The kinks repel, so they can approach
slowly from infinity, instantaneously stop at a large 
separation, and move out again to infinity. The separation remains large
throughout, so the effective equation of motion should be reliable. 
To test this picture, it would be helpful to perform a numerical 
simulation of the kink dynamics. Very recently, Belendryasova and Gani
\cite{BG} have studied kink-antikink dynamics numerically in the
same model as that discussed below, surprisingly finding among other 
things that at large separation and slow speeds, the kink and antikink repel.   

\vspace{7mm}

\section{A Model for a Kink with a Long-range Tail}
\vspace{4mm}

There are an unlimited number of scalar field theories with kinks having
long-range tails. Several of them, arising from a variety of polynomial
potentials $V$, are discussed in refs.\cite{Loh,KCS,BMM} and 
elsewhere. However, in almost all cases, the kink solution is 
not explicit, so the algebra needed to
investigate the kink's properties is complicated. We shall therefore
focus on just the simplest symmetric potential that admits two
related kinks with long-range tails, and calculate the force between
these. Our results should generalise.

The potential we consider is the octic polynomial
\be
V(\phi) = \half (1 - \phi^2)^2 \phi^4 \,.
\ee
This has quadratic minima at $\phi = \pm 1$, and a quartic minimum at
$\phi = 0$. The kink interpolates between $\phi = 0$ and $\phi =
1$, and there is a mirror kink that interpolates between $\phi = -1$
and $\phi = 0$, with the same energy.

The potential can be expressed as 
\be
V = \half \left(\frac{dW}{d\phi}\right)^2
\ee
where
\be
\frac{dW}{d\phi} = (1 - \phi^2)\phi^2 \,,
\ee
so
\be
W(\phi) = \frac{1}{3}\phi^3 - \frac{1}{5}\phi^5 + {\rm const.}
\ee
The constant has no significance, but it will be convenient to
set it to $-\frac{2}{15}$, so that $W(1) = 0$ and $W(0) = -\frac{2}{15}$.  

Starting with the Lagrangian density (\ref{Lagran}), and writing $V$ 
in terms of $W$, we obtain the energy expression for a static field
\be
E = \int_{-\infty}^{\infty} \left( \half\left(\frac{d\phi}{dx}\right)^2
+  \half\left(\frac{dW}{d\phi}\right)^2 \right) dx \,.
\ee
Using the Bogomolny rearrangement \cite{Bo} (completing the square and
integrating the cross term) we can reexpress this as
\be
E = \int_{-\infty}^{\infty} \half\left( \frac{d\phi}{dx}
-  \frac{dW}{d\phi} \right)^2 dx + W(1) - W(0) \,,
\ee
where we are assuming that the field interpolates between $\phi = 0$ as
$x \to -\infty$ and $\phi = 1$ as $x \to \infty$. The kink minimises
the energy (for the given boundary conditions) by satisfying the
Bogomolny equation
\be
\frac{d\phi}{dx} = \frac{dW}{d\phi} = (1 - \phi^2)\phi^2 \,,
\label{Bogoeq}
\ee
and its energy is 
\be
E = W(1) - W(0) = \frac{2}{15} \,.
\ee
The Bogomolny equation can be rewritten as
\be
\left( \frac{1}{2(1 - \phi)} + \frac{1}{2(1 + \phi)} +
  \frac{1}{\phi^2} \right)d\phi = dx \,,
\ee
whose implicit solution is \cite{Loh}
\be
\half\log\frac{1+\phi}{1-\phi} - \frac{1}{\phi} = x - A \,,
\label{exactkink}
\ee
where $A$ is a constant parameter. Inverting, we obtain the kink
solution $\phi(x-A)$. We call $A$ the location of the 
kink. Another interesting position in the kink is where 
the potential $V$ has its maximum value. This is where 
$\phi = \frac{1}{\sqrt{2}}$, because $\frac{d^2W}{d\phi^2}$ is 
zero here. We call this the centre of the kink. In the present 
example it occurs at $x_{\rm centre} = A + \log(1 + \sqrt{2}) - \sqrt{2} 
\simeq A - 0.533$. This is rather awkward to deal with algebraically, 
so we usually work with the location $A$. We further clarify the notion of
kink centre in Appendix A. Nothing physical depends on the 
distinction between location and centre.  

We need to know the tail behaviours of the kink. On the left, where 
$\phi$ is near zero, the Bogomolny equation simplifies to 
$\frac{d\phi}{dx} = \phi^2$, with tail solution
\be
\phi = \frac{1}{A - x} \,.
\ee
$A$ is the same constant as before. The next term in
the expansion of $\phi$ is cubic in $\frac{1}{A - x}$, which makes the
constant $A$ in the leading term unambiguous. $A$
is the location where the extrapolated tail field diverges. 
On the right, where $\phi = 1 - \eta$ with $\eta$ small and 
positive, the Bogomolny equation linearises to $\frac{d\eta}{dx} =
-2\eta$, with tail solution $\eta = \exp{-2(x-b)}$. The constant $b$ equals
$A-1+\half\log 2 \simeq A - 0.653$. Our main interest will be in the 
long-range tail on the left.

The transition between the tail on the left and the asymptotic field
value on the right is rather fast (see Fig. 4 in \cite{KCS}, or Fig. 2
in \cite{BG}), so a crude approximation to the kink with location 
$A=0$ is $\phi = -\frac{1}{x}$ for $x \le -1$ and $\phi = 1$ for $x \ge -1$. 

In the Lorentz invariant theory we are considering, the energy $E$ of a
kink is the same as its rest mass $M$. $M$ is of course the conversion
factor between force and acceleration for kinks moving
non-relativistically. We will also need to consider the energy (mass) of a kink
with its long-range tail truncated. Consider therefore the kink
with location $A$ on the half-line $X \le x < \infty$, with $X \ll
A$. Using the Bogomolny rearrangement again, we estimate that the tail
truncation reduces the energy by
\be
E_{\rm tail} = W(\phi(X)) - W(0) = \frac{1}{3(A-X)^3} \,,
\label{tailenergy}
\ee
where we have just retained the leading term in the kink solution, and in
$W$.

We conclude this section with some remarks about the well-known
mechanical reinterpretation of a kink solution. The equation for a
static field, obtained from the Lagrangian density (\ref{Lagran}), is
\be
\frac{d^2\phi}{dx^2} = \frac{dV}{d\phi} \,.
\ee
This can be interpreted as the Newtonian equation of motion for a unit mass
particle with ``position'' $\phi$ moving in ``time'' $x$ in the
inverted ``potential'' $-V$. For our kink, the particle falls
off the ``potential'' maximum at $\phi = 0$ and 
eventually stops at the  ``potential'' maximum at $\phi = 1$. Because the
total ``energy'' is zero, the motion also obeys the first-order 
Bogomolny equation. The motion away from the maximum at $\phi = 0$ is
particularly slow, because that maximum is quartic, leading to
the long-range kink tail, but the approach to the maximum at $\phi =
1$ occurs more rapidly. Below, we will consider an accelerating kink
(in the true time $t$) and will find that it approximately satisfies a
modified static equation. In the mechanical reinterpretation, this is
a Newtonian equation of motion with friction, so to get a solution for which 
the particle eventually stops at $\phi = 1$, the particle needs to leave
$\phi = 0$ with a positive ``velocity'' at a finite ``time''. We shall
make the approximation that the friction is operative mainly during
the long, slow ``descent'' of the particle away from $\phi = 0$, and is
negligible during the rapid ``ascent'' to $\phi = 1$. The frictional
``force'' is comparable during these stages, since the ``velocities''
are similar, but the ``times'' over which it acts are very different. 

\vspace{7mm}

\section{Force between Kinks using a Static Interpolating Field}
\vspace{4mm}

In addition to the kink solution, our model has a
mirror kink that interpolates between $\phi = -1$
and $\phi = 0$. If $\phi(x)$ is the kink, then $-\phi(-x)$ is the mirror
kink. Note that the mirror kink obeys the same Bogomolny equation as
the kink,
\be
\frac{d\phi}{dx} = \frac{dW}{d\phi} \,.
\ee
This is rather unusual, and is a consequence of $W$ having a cubic
stationary point at $\phi = 0$. The equation still has no static solution 
with both a mirror kink and kink, interpolating between $\phi = -1$ 
and $\phi = 1$. This is because the Bogomolny equation is a gradient flow
equation for the superpotential $W$, and solutions cannot pass through
stationary points of $W$. Instead, the mirror kink and kink repel
each other, and we shall calculate their accelerations when the kink
is located at $A$ and the mirror kink is at $-A$, for $A \gg 0$. 
We may assume the field is antisymmetric in $x$ at all times.

We need to produce a sensible interpolating field between 
$\phi = -1$ and $\phi = 1$. Simply adding the mirror kink and kink 
fields is not a good idea, even though the desired boundary conditions 
are satisfied. This is because the long-range tail of the kink would extend 
right across the mirror kink, destroying its short-range tail and 
producing the wrong asymptotic field as $x \to -\infty$, and similarly
the long-range tail of the mirror kink would extend right across the kink.

Instead we split up the spatial line at points $-X$ and $X$, with $0 \ll
X \ll A$, so that for $x \le -X$ we have an exact mirror kink field, and
for $x \ge X$ an exact kink field. In between, for $-X \le x \le X$, we 
assume the interpolating field has the linear behaviour $\phi(x) = \mu
x$. This is justified by the linearised, static field equation for small
$\phi$, which is simply $\frac{d^2\phi}{dx^2} = 0$.   
We require the field and its first spatial derivative to be continuous
at $X$ (and also $-X$). This leads to the conditions
\be
X = \half A \quad {\rm and} \quad \mu = \frac{4}{A^2} \,,
\label{Xmu}
\ee
where we have assumed the tail formula for the kink field, which 
is approximately valid at $X$.

Note that if we had solved the static field equation exactly between
the mirror kink and kink, the appropriate continuity conditions would 
have been impossible to solve, as they would have given us a global 
smooth, static solution, which doesn't exist. This impasse is resolved by
using an approximate solution, as we have done.

Let us now calculate the total energy of this interpolating field, to
leading order in $\frac{1}{A}$. This is the sum of the energies of the
kink and mirror kink, with their tails truncated, and the energy of the linear
interpolating field. We calculated the tail energy in
eq.(\ref{tailenergy}). For the linear part of the field, the energy is
\be
E = \int_{-X}^{X} \left( \half\left(\frac{d\phi}{dx}\right)^2
+  \half\left(\frac{dW}{d\phi}\right)^2 \right) dx \,.
\ee 
Here we set $\phi = \mu x$ and make the approximation
$\frac{dW}{d\phi} = \phi^2 = \mu^2 x^2$. The integral gives $\mu^2 X +
\frac{1}{5} \mu^4 X^5$. Now we use the values (\ref{Xmu}) and find that
$\mu^2 X$ and $\mu^4 X^5$ are both of the same order in $\frac{1}{A}$
as the tail energy, and the total energy is
\be
E = \frac{4}{15} - \frac{16}{3 A^3} + \frac{8}{A^3} + \frac{8}{5 A^3}
= \frac{4}{15} + \frac{64}{15 A^3} \,.
\ee 
Using this as the potential energy, and dropping the constant, we
obtain an effective Lagrangian for the motion of the two kinks
\be
L_{\rm kinks} = \frac{2}{15} {\dot A}^2 - \frac{64}{15 A^3} \,,
\ee
where the kinetic energy is twice 
that of a single kink. The equation of motion is
\be
{\ddot A} = \frac{48}{A^4} \,,
\label{accel1}
\ee
showing that the kink has acceleration $\frac{48}{A^4}$ to the
right. The mirror kink has opposite acceleration, so the kink and 
mirror kink repel each other. The force acting is the mass 
$M = \frac{2}{15}$ times the acceleration, and is   
\be
F = \frac{32}{5 A^4} \,.
\ee
The separation of kink and mirror kink is $2A$ with an uncertainty of
order $1$, but this uncertainty does not affect the force to leading
order in $\frac{1}{A}$.

This calculation gives, we believe, the correct dependence on $A$, but
the coefficient is not correct. This is because our result is
sensitive to the interpolation used. Also, the force is not acting
uniformly throughout each kink, but only in the region of the linear
interpolating field. One knows this because in the regions of the
exact, static kink and mirror kink solutions, no force 
is acting. In the next section we will use a different approach, where 
the kinks are actually accelerating, and deformed, and will 
derive a more reliable force. 

\vspace{7mm}

\section{Accelerating Kinks}
\vspace{4mm}

In this second approach to the force between the kink and mirror kink,
we model the time-dependent field $\phi(x,t)$, assuming
that the kink and mirror kink are located at time-dependent positions 
$A(t)$ and $-A(t)$, and are well separated.  

The idea is to find an implicit profile for the kink with acceleration
$a = {\ddot A}$ that describes the field for $x>0$, and at least
approximately solves the field equation. The mirror kink has
the reflected field (sign-reversed $x$, $\phi$ and $a$) for 
$x<0$. The accelerating kink has a distorted profile, whose tail
is no longer $\frac{1}{A-x}$. Instead, the field $\phi$ is zero at some
finite distance to the left of the kink, the distance depending on the 
acceleration. The kink profile is linear near here, so it can be 
continuously glued to the profile of the accelerating mirror kink. The
gluing can be carried out at $x=0$ to create a field antisymmetric in
$x$, with a continuous first derivative.

This interpolation does not give an exact solution, because the
acceleration jumps discontinuously from $a$ to $-a$ at $x=0$, and
because we need to make various approximations that are explained in
more detail below. Tail gluing nevertheless creates a convenient interpolation.

We recall here a result related to Noether's theorem for
conserved momentum. By a standard argument \cite{book}, the momentum
density in the field theory we are considering is $-\frac{\pr\phi}{\pr
  t}\frac{\pr\phi}{\pr x}$, and its integral over the whole spatial line  
is conserved provided $\phi$ satisfies its field equation. The time
derivative of the momentum integral over a finite interval is a sum of endpoint 
(surface) terms. In particular, using the field equation, one finds that 
for a field on the half-line $x \ge X$, obeying the kink boundary 
condition as $x \to \infty$, and for which 
$\left(\frac{\pr\phi}{\pr t}\right)^2$ is negligible, 
\be
\frac{d}{dt} \int_X^\infty -\frac{\pr\phi}{\pr t}
\frac{\pr\phi}{\pr x} \, dx = \left(\half\left(\frac{\pr\phi}{\pr x}\right)^2 
- V(\phi(x))\right) \Bigg|_{x=X} \,.
\label{force}
\ee
One may interpret the rate of change of momentum, given here, as the 
force $F$ acting on the entire field to the right of $X$, and it
appears to be exerted at the point $X$, although we shall refine 
this interpretation later. For the field of the accelerating kink and
mirror kink we can choose $X=0$, and then $\phi = 0$ and $V = 0$. The
force on the kink becomes
\be
F = \half\left(\frac{\pr\phi}{\pr x}\right)^2 \Bigg|_{x=0} \,.
\label{Forceslope}
\ee

The last formula can be used to get a rough estimate for the force,
even when one does not attempt to solve the field equation. For
example, if we simply add the static tail fields of the kink and 
mirror kink, then near $x=0$,
\be
\phi \simeq \frac{1}{A-x} - \frac{1}{A+x} \,.
\label{tailsum}
\ee
At $x=0$, we find $\frac{d\phi}{dx} = \frac{2}{A^2}$, so the force on
the kink is estimated to be $F = \frac{2}{A^4}$, and its acceleration
${\ddot A} = \frac{15}{A^4}$. This has the right
quartic dependence on $\frac{1}{A}$, but the coefficient does not
agree with what we found earlier, nor with what we claim is the more
accurate value to be determined later in this section. 

Now let's return to finding an approximate solution of the field equation
representing the accelerating kink. We do not need to
know the cause of the acceleration, but it will in fact be produced by
the mirror kink to the left. We model the kink by a field of the form
\be
\phi(x,t) = \chi(x - A(t)) \,.
\label{accelfield}
\ee
The acceleration is $a = {\ddot A}$ and we assume this is small. We
also assume the squared velocity ${\dot A}^2$ remains much less than $1$, so the
motion is non-relativistic, and we can ignore Lorentz contraction of
the kink, and radiation. Throughout, we work to linear order in $a$. 
It is convenient to denote the argument of $\chi$ by $y$ and denote 
a derivative of $\chi$ with respect to $y$ by $'$.

The first time derivative of $\phi$ is $-{\dot A}\chi'$, and we approximate   
the second time derivative by $-{\ddot A}\chi' = -a\chi'$, dropping a term
proportional to ${\dot A}^2$. The dropped term is negligible while the motion
remains non-relativistic. (The term $\half\left(\frac{\pr\phi}{\pr t}\right)^2$ 
that would be on the right hand side of eq.(\ref{force}) is similarly
proportional to ${\dot A}^2$, which justifies its neglect.) 
Substituting the accelerating field (\ref{accelfield}) into the full 
field equation (\ref{fieldeq}) then gives
\be
\chi'' + a\chi' - \frac{dV(\chi)}{d\chi} = 0 \,.
\label{chieq}
\ee
The profile $\chi$ satisfies this static equation with $a$ as parameter, 
so it evolves adiabatically with time as $a$ varies. $\phi$ evolves both
because of this evolution of $\chi$, and because $A(t)$ occurs in the 
argument of $\chi$. 

Using the mechanical analogy discussed in Section 2, eq.(\ref{chieq}) is
interpreted as the motion of a ``particle'' with ``position'' $\chi$ in the 
inverted potential $-V$, now subject to friction with a friction 
coefficient $a$.
The kink boundary condition is that $\chi$ approaches $1$ as $y \to \infty$.
The potential $-V$ is zero at both $\chi = 0$ and $\chi = 1$, so for the
``particle'' to reach $\chi = 1$ eventually and stop there, it must leave
$\chi = 0$ with a finite ``velocity'' at a finite, but arbitrary
time. This ``velocity'' is determined by $a$. 

Reverting back to field language, we anticipate a solution 
of eq.(\ref{chieq}) where $\chi = 0$ for some finite $y$, 
and $\chi \to 1$ as $y \to \infty$. The equation for $\chi$ is
translation invariant, so any solution can be shifted to the left or
right. The solution we require is the one where $\chi(x-A)$ agrees as
closely as possible with the static kink solution $\phi(x-A)$ in the core
region of the kink. 

This requirement is however not so easy to implement. The short-range 
tails to the right have slightly different exponential forms, so their
coefficients cannot really be matched. Instead, one could require, for example,
that $\phi$ and $\chi$ take identical values $\frac{1}{\sqrt{2}}$ at
the same position. This could be implemented numerically, but it
cannot be done analytically because we do not have even an implicit
solution for $\chi$. 

The approach we have adopted to this problem is to match the long-range 
tails of $\phi$ and $\chi$ close to the core regions, and 
in particular to arrange that the positions where the extrapolated 
long-range tails diverge are the same. This assumes that the main 
effect of the friction term is on these tails, and its effect in the 
kink core and further to the right is negligible, as we argued earlier.

For the static kink, the extrapolated tail $\frac{1}{A-x}$ diverges at $x =
A$. For $\chi$ we calculate as follows. In the small term $a\chi'$ in 
eq.(\ref{chieq}) we can assume that $\chi$ is the undeformed kink, for 
which $\chi' = \frac{dW}{d\chi}$. We can therefore trade the friction 
term for a modified potential $\widetilde{V} = V - aW$ and obtain a 
first integral of eq.(\ref{chieq}) of the form
\be
\chi'^2 = 2\widetilde{V} + {\rm const.}
\label{modeqchi}
\ee
The constant is zero, because $\chi'$, $V$ and $W$ are all zero as 
$y \to \infty$. In the long-range tail region we can now make the 
approximations $V(\chi) = \half\chi^4$ and $aW = -\frac{2a}{15}$, 
obtaining from eq.(\ref{modeqchi}) the simplified equation
\be
\frac{d\chi}{dy} = \sqrt{\frac{4a}{15} + \chi^4} \,.
\label{simplechi}
\ee
As a consistency check, note that the linear behaviour of $\chi$ near
its zero has slope $\mu = \sqrt{\frac{4a}{15}}$, so the force on the
kink, according to eq.(\ref{force}), is $\frac{2a}{15}$. This is the
product of the kink's mass and acceleration, as it should be.  

The solution of eq.(\ref{simplechi}) still involves an
elliptic integral of the first kind, but fortunately we just need the definite
integral. The solution $\chi(y)$ should have the properties $\chi(-A)
= 0$ and $\chi(0)$ diverges. Then $\chi(x-A)$ will be zero at
$x=0$ and $\chi(x-A)$ will diverge at $x=A$, as we require for the 
extrapolated tail field. Therefore
\be
\int_0^\infty \frac{d\chi}{\sqrt{\frac{4a}{15} + \chi^4}} = A \,.
\ee
This complete elliptic integral is known \cite{GR}, and gives
\be
A = \left(\frac{4a}{15}\right)^{-\frac{1}{4}}
\frac{\Gamma\left(\frac{1}{4}\right)^2}{4\sqrt{\pi}} \,.
\ee
Inverting, we find
\be
a = {\ddot A} = \frac{15}{4} 
\left(\frac{\Gamma\left(\frac{1}{4}\right)^2}{4\sqrt{\pi}}\right)^4
\frac{1}{A^4} \simeq \frac{15}{4}(1.854)^4 \frac{1}{A^4} =
\frac{44.3}{A^4} \,.
\ee
This is our result for the acceleration. The coefficient $44.3$ is
different from the previous estimates of $48$ and $15$. The force 
on the kink is
\be
F = \frac{2}{15}{\ddot A} \simeq \half(1.854)^4 \frac{1}{A^4}
= \frac{5.91}{A^4} \,.
\ee
We can go back through the calculation, and verify that terms that were
dropped, e.g. $\frac{1}{3}a\chi^3$ in $aW$, are small compared to 
those we retained. 

The equation of motion for the kink, ${\ddot A} = \frac{44.3}{A^4}$,
implies the energy conservation equation 
\be
\half {\dot A}^2 + \frac{1}{3}\frac{44.3}{A^3} = {\rm const.}
\ee
and from this we deduce that if the kinks approach from infinity with
a small speed $v$, then at closest approach
\be
A = \left(\frac{29.5}{v^2}\right)^{\frac{1}{3}} \,.
\ee

This calculation, involving an accelerating kink, seems to give a
more reliable result than our earlier, static approach because 
the force is properly distributed along the kink. 
To see this, we multiply eq.(\ref{chieq}) by $\chi'$ and integrate
from $X$ to $\infty$, for $X \ge 0$, obtaining
\be
a\int_X^{\infty} \chi'^2 \, dy = \left( \half\chi'^2 -
  V(\chi(y))\right) \Bigg|_{y=X} \,.
\ee
By analogy with eq.(\ref{force}), the right hand side is the force
acting on the half-line to the right of $X$. Since $a$ is small, on
the left hand side we can replace $\chi'$ by the derivative of the
static kink. The Bogomolny equation then implies that $\chi'^2$ is the 
energy density of the static kink, to the accuracy we need. (Of course, we
can't use the Bogomolny equation on the right hand side, because this 
would give zero.) Therefore the left hand side is the mass on the 
half-line to the right of $X$, times the acceleration 
$a$. As $X$ is not fixed, we deduce that the force is everywhere of 
the correct strength.

We have apparently lost track of the mirror kink, but this occupies
the half-line $-\infty < x \le 0$, and its field is
$-\phi(-x,t) = -\chi(-x-A(t))$. The 
acceleration of the mirror kink is $-a$. The kink and mirror kink
fields meet at $x=0$ and have a continuous derivative. The mirror kink
satisfies an equation like (\ref{chieq}) but with the sign of $a$
reversed. The second spatial derivative of the field $\phi$ at $x=0$ 
therefore has a discontinuity $2a\chi'$, which we see from 
eq.(\ref{simplechi}) is of order $a^{\frac{3}{2}}$. This discontinuity
is presumably negligible, but is a consequence of the several 
approximations we have made. An even better approximation
would involve a smoother interpolation of the acceleration through $x=0$.

\vspace{7mm}

\section{Conclusions}
\vspace{4mm}

We have investigated a simple example of a kink with a long-range
tail in scalar field theory. The tail has $\frac{1}{x}$ behaviour, 
because the potential $V$ in the field theory has a quartic minimum. We 
have considered the force between two such kinks, when their long-range 
tails overlap in the region between them. The force is proportional to
the inverse fourth power of the kink separation when the separation is 
large, and is repulsive. The numerical coefficient has been calculated 
by allowing for the kink accelerations and then solving a modified 
Bogomolny equation, and is 
transcendental. It should not be difficult to generalise our discussion 
to kink-antikink forces, and to variant kinks with long-range
tails. It would be interesting to solve the equation of motion for
incoming kinks with small velocity, using the force we have obtained,
and compare with a numerical simulation of kink dynamics using the 
full field equation.

\vspace{.5cm}

%%%%%%%%% Acknowledgements %%%%%%%%%%%%%
\section*{Acknowledgements}
%%%%%%%%%%%%%%%%%%%%%%%%%%%%%%%%%%%%%%%%

I am grateful to Avadh Saxena for drawing my attention to this
problem, and for subsequent correspondence. I also thank Avadh, Emil 
Mottola, and their colleagues, for hospitality during my visit to Los 
Alamos in August 2018. This work has been partially supported by 
STFC consolidated grant ST/P000681/1.

\vspace{7mm}

\section*{Appendix A: Kink Centre}
\vspace{4mm}

Let $\phi_n$ and $\phi_{n+1}$ be adjacent zeros of $\frac{dW}{d\phi}$,
and hence of the potential $V = \half\left(\frac{dW}{d\phi}\right)^2$. 
We assume $\frac{dW}{d\phi}$ is positive in between these zeros. Then
the Bogomolny equation $\frac{d\phi}{dx} =\frac{dW}{d\phi}$ has the 
correct sign for a kink interpolating between $\phi_n$ and
$\phi_{n+1}$, and the kink solution $\phi(x)$ increases 
monotonically with $x$.

$V$ could have several local maxima and minima between $\phi_n$ and
$\phi_{n+1}$, but let us assume that it has just one maximum, denoted 
by $\phi_{n+\half}$. For the following reasons, the position 
$x = x_{\rm centre}$ where $\phi = \phi_{n+\half}$ can be regarded 
as the centre of the kink. At $x_{\rm centre}$, 
\be
\frac{dV}{d\phi} = \frac{dW}{d\phi}\frac{d^2W}{d\phi^2} = 0 \,,
\ee
so $\frac{d^2W}{d\phi^2} = 0$, as $\frac{dW}{d\phi}$ is
non-zero. Differentiating the Bogomolny equation gives, at $x_{\rm centre}$,
\be
\frac{d^2\phi}{dx^2} = \frac{d^2W}{d\phi^2} \frac{d\phi}{dx} = 0 \,,
\ee
so $x_{\rm centre}$ is the point of inflection in the kink
profile. It is the position where $\phi$ is increasing with $x$ most
rapidly. The energy density of a kink satisfying the Bogomolny
equation is $2V$, so this is also maximal at $x_{\rm centre}$.
Let $\phi(x)$ be a centred kink, for which $x_{\rm centre} = 0$. The 
general kink is then $\phi(x-c)$ with centre $c$. 

For the familiar $\phi^4$ and sine-Gordon kinks, the
centres are the obvious points about which the kink is antisymmetric.
The centre of the kink with long-range tail, considered in the main
part of this paper, was clarified in Section 2.
Another non-trivial example is the $\phi^6$ kink \cite{Loh}. Here,
\be
V(\phi) = \half(1-\phi^2)^2\phi^2 \,,
\ee
with quadratic minima at $\pm 1$ and $0$. The quadratic behaviours are
different at $0$ and $1$, so the interpolating kink solution is not 
reflection (anti)symmetric. As 
$\frac{dW}{d\phi} = (1-\phi^2)\phi$, 
\be
W(\phi) = \half \phi^2 - \quart \phi^4 + {\rm const.}
\ee
and the Bogomolny equation is 
\be
\frac{d\phi}{dx} = (1-\phi^2)\phi \,.
\ee
Using partial fractions, this can be integrated to give
\be
\phi(x) = \left(1 + 2e^{-2(x-c)}\right)^{-\half} \,.
\label{phi6kink}
\ee   
The kink centre is where $\frac{d^2W}{d\phi^2} = 1 - 3\phi^2 = 0$, 
i.e. where $\phi = \frac{1}{\sqrt{3}}$. The expression (\ref{phi6kink}) has 
been carefully normalised so that $x_{\rm centre}=c$. The kink energy 
is $E = W(1) - W(0) = \quart$.  

\vspace{7mm}

\section*{Appendix B: Force between $\phi^6$ Kinks}
\vspace{4mm}

Here, for completeness, we illustrate the simple method that 
gives the force between two kinks having short-range tails. The 
following example was not explicitly considered in \cite{Ma5,book}.

The $\phi^6$ theory has kink solution (\ref{phi6kink}) interpolating
between $\phi = 0$ and $\phi = 1$, and a mirror kink interpolating
between $\phi = -1$ and $\phi = 0$ obtained by reversing the signs 
of $x$ and $\phi$. A field with a kink centred at $c$
and mirror kink centred at $-c$, with $c \gg 0$, is well described 
by the linear superposition
\be
\phi(x) = \left(1 + 2e^{-2(x-c)}\right)^{-\half} - 
\left(1 + 2e^{2(x+c)}\right)^{-\half} \,. 
\label{phi6super}
\ee
The kinks tails are short-ranged here, i.e. exponentially decaying, so
the linear superposition is better justified than for the kinks with
long-range tails we discussed earlier. The kink obeys the Bogomolny
equation and the mirror kink the equation with reversed sign. So the 
linear superposition does not obey either Bogomolny equation, and is 
therefore not an exact static solution. However, both kink tails obey 
the linearised second-order static field equation in the region
between the kinks, so the sum of the tails does too. This explains why the
linear superposition is a good interpolating field.

To find the force exerted by the mirror kink on the kink, we use the
Noether formula for the rate of change of momentum 
(\ref{force}). At a point $X$ between the kinks, with $-c \ll X \ll
c$, the force acting on the kink to the right is
\be
F = \left(\half\left(\frac{d\phi}{dx}\right)^2 - V(\phi)\right)
\Bigg|_{x=X} \,.
\ee
We can make the approximation $V(\phi) = \half\phi^2$ here, as $\phi$
is near zero, so
\be
F \simeq \left(\half\left(\frac{d\phi}{dx}\right)^2 - \half\phi^2 \right)
\Bigg|_{x=X} \,.
\label{tailsforce}
\ee
The superposed tail field, obtained from the leading exponentially
small terms in (\ref{phi6super}), is
\be
\phi(x) = \frac{1}{\sqrt{2}}e^{x-c} - \frac{1}{\sqrt{2}}e^{-x-c} \,.
\ee
Each term separately would give no force, so it is the cross
terms that produce a non-zero result. Substituting into 
(\ref{tailsforce}), we find that the force is
\be
F = e^{-2c} \,,
\ee
independent of $X$. As the mass of the kink is $\quart$,
its acceleration is $4e^{-2c}$, and the mirror kink has the opposite
acceleration. Because the kink tail has exponentially small energy, 
the kink mass to the right of $X$ is effectively constant even as 
$X$ varies, and therefore an $X$-independent force is what's needed 
to produce a definite acceleration.  

The effective equation of motion for the kink is 
\be
\quart {\ddot c} = e^{-2c} \,,
\label{cmotion}
\ee
implying conservation of energy
\be
\quart {\dot c}^2 + e^{-2c} = {\rm const.}
\ee
From this one can determine the closest approach, $c = \log 2 - \log v$, 
if the kink and mirror kink approach from infinity with speed $v$. 

Note that if we had decided on a different notion of the kink 
location, say $A = c + \delta$, with $\delta$ of order $1$, we would 
have derived the equation of motion
\be
\quart {\ddot A} = e^{2\delta}e^{-2A} \,.
\label{Amotion}
\ee 
This shows that the coefficient in front of an exponentially small
force is only meaningful if one is very careful to specify where the 
kink is located. Equations (\ref{Amotion}) and (\ref{cmotion}) are 
completely equivalent, although they seem to predict a closest
approach for the kinks differing by $\delta$.

\end{document}